\def\kF{k_{\text{F}}}

\def\NF{N_{\text{F}}}

\def\epsilonF{\epsilon_{\text F}}

\def\sgn{{\text{sgn\,}}}
\def\be{\begin{equation}}
\def\ee{\end{equation}}
\def\bea{\begin{eqnarray}}
\def\eea{\end{eqnarray}}
\def\bse{\begin{subequations}}
\def\ese{\end{subequations}}

\documentclass[prb,twocolumn,amsmath,amssymb,eqsecnum,preprintnumbers]{revtex4}
\usepackage{graphicx}
\usepackage{dcolumn}
\usepackage{bm}
\usepackage{bbm}
\usepackage{color}
\begin{document}
\title{Exact solution of the Boltzmann equation for low-temperature transport coefficients in metals I: 
         Scattering by phonons, antiferromagnons, and helimagnons}
\author{J. Amarel$^1$, D. Belitz$^{1,2}$, and T.R. Kirkpatrick$^3$}
\affiliation{$^{1}$ Department of Physics and Institute for Fundamental Science,
                    University of Oregon, Eugene, OR 97403, USA\\
                   $^{2}$ Materials Science Institute, University of Oregon, Eugene,
                    OR 97403, USA\\
                  $^{3}$ Institute for Physical Science and Technology,
                    University of Maryland, College Park,
                    MD 20742, USA
            }
\date{\today}

\begin{abstract}
We present a technique for an exact solution of the linearized Boltzmann equation for the electrical and thermal 
transport coefficients in metals in the low-temperature limit. This renders unnecessary an uncontrolled
approximation that has been used in all previous solutions of the integral equations for the transport coefficients.
Applications include electron-phonon scattering in nonmagnetic metals, as well as the magnon contribution to the 
electrical and thermal conductivities, and to the thermopower, in metallic ferromagnets, antiferromagnets, and 
helimagnets. In this paper, the first of a pair, we set up the technique and apply it to the scattering of
electrons by phonons, antiferromagnons, and helimagnons. We show that the Bloch $T^5$ law for the electrical resistivity, the
$T^2$ law for the thermal resistivity, and the $T$ law for the thermopower due to phonon and antiferromagnon scattering
are exact, and determine the prefactors exactly. The corresponding exact results for helimagnons are $T^{5/2}$, $T^{1/2}$,
and $T$, respectively. In a second paper we will consider the scattering by ferromagnons.
\end{abstract}


\maketitle

\section{Introduction}
\label{sec:I}

It is well known that collective excitations that are soft, or massless, at zero temperature
($T=0$) lead to power-law behavior of transport coefficients in the limit $T\to 0$. The
resulting power depends on the frequency-momentum relation of the soft mode, and on
its coupling to the conduction electrons. An old example is Bloch's $T^5$ law for the electrical 
resistivity $\rho$ due to scattering by acoustic phonons.\cite{Bloch_1930, Ziman_1960} The analog
of Bloch's law for magnon scattering in metallic ferromagnets was investigated by Ueda and Moriya,\cite{Ueda_Moriya_1975}
who found a $T^2$ contribution to the electrical resistivity. Other examples include scattering
by magnons in antiferromagnets, which yield a $T^5$ contributions as phonons do,\cite{Yamada_Takada_1974, Ueda_1977}
and in helimagnets, where the corresponding behavior is $T^{5/2}$.\cite{Belitz_Kirkpatrick_Rosch_2006b}

All of these results were obtained by solving either the linearized Boltzmann equation,
or an equivalent integral equation derived from the Kubo formula, and the solutions involved an uncontrolled 
approximation that replaces various energy-dependent relaxation rates by constants. The results hence were
not as well founded as is desirable, and on occasion even the validity of the Bloch $T^5$ law has
been doubted, see Ref.~\onlinecite{Mahan_2000}. Only very recently has it been shown that a mathematically rigorous solution of the
integral equation does indeed yield the Bloch $T^5$ law for the phonon contribution to the electrical 
resistivity $\rho \propto T^5$.\cite{Amarel_Belitz_Kirkpatrick_2020} 

The purpose of the present paper is four-fold:
First, we show that the method of Ref.~\onlinecite{Amarel_Belitz_Kirkpatrick_2020} can be simplified
substantially by making some assumptions about the collision operator that are common in theoretical
physics, chiefly, the existence of a spectral representation. Second, we generalize the method to allow 
for a calculation of the thermopower $S$ and the heat conductivity $\sigma_h$ in addition to the electrical 
conductivity, and we establish that the leading temperature dependences, $S\propto T$ and
$\sigma_h \propto 1/T^2$, are exact. Third, we show that the scattering of electrons by antiferromagnons 
leads to the same temperature dependence of the transport coefficients as the scattering by phonons. 
Fourth, we calculate the prefactors of the power laws exactly. The result for the electrical conductivity is
a Drude formula,
\bse
\label{eqs:1.1}
\be
\sigma = n e^2\tau_{\sigma}/m\ ,
\label{eq:1.1a}
\ee
with $n$, $e$, and $m$ the electron density, charge, and effective mass, respectively, and a relaxation time
\be
\tau_{\sigma}(T\to 0) = \frac{1}{120\,\zeta(5) g_0}\,\frac{1}{1 + T_1^2/4\epsilonF^2}\,\frac{T_1^4}{T^5} \ .
\label{eq:1.1b}
\ee
\ese
Here $\epsilonF$ is the Fermi energy, and $T_1$ is the bosonic energy scale, which is on the order
of the Debye temperature for phonons, and its magnetic analog for antiferromagnons. $\zeta$ is the Riemann
zeta function, and $g_0$ is a dimensionless coupling constant that depends on the parameters of the electron-phonon or
electron-magnon coupling. The result for the thermopower is
\be
S(T\to0) = \frac{-\pi^2}{6\,e}\,\frac{T}{\epsilonF}\ .
\label{eq:1.2}
\ee
The result for the heat conductivity is also exact, but the prefactor involves an integral over a
scaling function that we have been unable to determine explicitly. 

Results for helimagnon scattering, which can be
derived in complete analogy to the phonon and antiferromagnon cases, are summarized in
an appendix. In a second paper\cite{Paper_II} (to be referred to as Paper II) 
we will analyze the scattering of electrons by ferromagnons, which is a more complicated problem.

This paper is organized as follows. In Sec.~\ref{sec:II} we recall the linearized Boltzmann equations
for the phonon or antiferromagnon scattering contributions to the electrical and thermal resistivities, 
as well as to the thermopower. 
In Sec.~\ref{sec:III} we present a method for solving the Boltzmann equations exactly
in the limit of asymptotically low temperature. 
Several aspects of our solution technique and our results are discussed in Sec.~\ref{sec:IV}. 
Some technical points related to the effective electron-electron interaction due to boson
exchange in general, and phonons and magnons in particular, are given in Appendix~\ref{app:A}.
Various relaxation rates are discussed in Appendix~\ref{app:B}. Appendix~\ref{app:C} sketches
the adaptation of our method for helimagnon scattering and gives the corresponding results, 
which also are exact.

\section{Transport coefficients, and the Boltzmann equation}
\label{sec:II}

\subsection{Transport coefficients}
\label{subsec:II.A}

Consider a mass current ${\bm J}$ and a heat current ${\bm J}_h$ driven by gradients of the electrochemical
potential $\bar\mu = \mu + eV$ and the temperature $T$, respectively. Here $\mu$ and $V$ are the chemical and electric potential,
respectively. To linear order in the potential gradients
the currents are determined by three independent transport coefficients (see, e.g., Ref.~\onlinecite{Mahan_2000}),
\bse
\label{eqs:2.1}
\bea
{\bm J} = -\frac{1}{T}\,L_{11} \bm\nabla\bar\mu - \frac{1}{T^2}\,L_{12}\,\bm\nabla T\ ,
\label{eq:2.1a}\\
{\bm J}_h = -\frac{1}{T}\,L_{12} \bm\nabla\bar\mu - \frac{1}{T^2}\,L_{22}\,\bm\nabla T\ .
\label{eq:2.1b}
\eea
\ese
Here we have used an Onsager relation that ensures that the two coefficients labeled $L_{12}$ are the same.
The electrical conductivity $\sigma$ is defined in the absence of a temperature
gradient, and for a constant chemical potential, via $e{\bm J} = -\sigma \bm\nabla V$, and the heat conductivity
$\sigma_h$ in the absence of an electrochemical potential gradient  via ${\bm J}_h = -\sigma_h\bm\nabla T$,
so
\bse
\label{eqs:2.2}
\be
\sigma = \frac{e^2}{T}\,L_{11}\quad , \quad \sigma_h = \frac{1}{T^2}\,L_{22}\ .
\label{eq:2.2a}
\ee
The Seebeck coefficient or thermopower $S$ is defined in the absence of a mass current via $\bm\nabla V = S\,\bm\nabla T$,
and hence
\be
-eS = \frac{1}{T}\,\frac{L_{12}}{L_{11}}\ .
\label{eq:2.2b}
\ee
Finally, the heat, or thermal, conductivity $\kappa$ in the absence of a mass current, which is the usual
experimental situation, is given by the combination
\be
\kappa = \frac{1}{T^2}\left(L_{22} - (L_{12})^2/L_{11}\right)\ .
\label{eq:2.2c}
\ee
\ese

The Onsager coefficients $L_{11}$ and $L_{22}$ are given by the Kubo expressions for the mass current - mass current
and heat current - heat current susceptibilities, respectively, and $L_{12}$ by the one for the mixed mass current -
heat current susceptibility.\cite{Kubo_1957, Mahan_2000} The Kubo formulas represent the exact linear response 
of the system and are very difficult to evaluate. They often are evaluated in a conserving approximation 
that is equivalent to the linearized Boltzmann equation,\cite{Wilson_1954, Mahan_2000} which we discuss next.

\subsection{The linearized Boltzmann equation}
\label{subsec:II.B}

\subsubsection{The linearized Boltzmann equation for the distribution function}
\label{subsubsec:II.B.1}

Let $f_0(x) = 1/(e^x+1)$ be the equilibrium Fermi-Dirac distribution, and $f$ the distribution in the presence of
an external perturbation. It is convenient to parameterize $f$ in terms of a function $\Phi({\bm k})$ by
\be
f({\bm k}) = f_0(\epsilon_{\bm k}/T) - \frac{\partial f_0(\epsilon_{\bm k}/T)}{\partial\epsilon_{\bm k}}\,\Phi({\bm k})\ ,
\label{eq:2.3}
\ee
where $\epsilon_{\bm k}$ is the electronic single-particle energy. For simplicity we will assume a parabolic band, 
$\epsilon_{\bm k} = {\bm k}^2/2m$, and in the main text of this paper we consider scattering processes that do not depend on the
electron spin in an essential way. This includes scattering by phonons, antiferromagnons, and helimagnons;
in Paper II we will generalize to the ferromagnetic case where the spin dependence is crucial. 
We are interested in perturbations that consist of an electric field ${\bm E} = -\bm\nabla V$ and a 
temperature gradient $\bm\nabla T$. The linearized Boltzmann equation then reads\cite{Wilson_1954, Ziman_1960}
\be
-e{\bm E}\cdot{\bm\nabla}_{\bm k} f({\bm k}) + \frac{1}{m}\,{\bm k}\cdot{\bm\nabla T}\,\frac{\partial f}{\partial T}= \left(\frac{\partial f}{\partial t}\right)_{\text{coll}}({\bm k})\ .
\label{eq:2.4}
\ee
It expresses the balance between the streaming term on the left-hand side and the temporal change of the
distribution function due to collisions on the right-hand side. To linear order in $\Phi$ the collision term can be
written\cite{Wilson_1954}
\bse
\label{eqs:2.5}
\bea
\left(\frac{\partial f}{\partial t}\right)_{\text{coll}}^{\text{lin}}({\bm k}) &\equiv& ({\mathfrak C}\Phi)({\bm k}) = 
\nonumber\\
&&\hskip -50pt = \frac{-1}{\kF^3} \int d{\bm k}'\,W({\bm k},{\bm k}')\,\left[\Phi({\bm k}) - \Phi({\bm k}')\right] \ . \hskip 20pt
\label{eq:2.5a}
\eea
with the kernel $W$ given by
\bea
W({\bm k},{\bm k}') &=& g_0\,\frac{T_1 \epsilonF}{T} \frac{\vert {\bm k}' - {\bm k}\vert^2}{2m} \frac{n_0(\omega_{{\bm k}'-{\bm k}}/T)}{\omega_{{\bm k}'-{\bm k}}}
\nonumber\\
&& \hskip -50pt  \times
     \left[f_0(\xi_{\bm k}/T)\left(1 - f_0(\xi_{{\bm k}'}/T)\right) \delta\left(\epsilon_{\bm k} - \epsilon_{{\bm k}'} + \omega_{{\bm k}'-{\bm k}}\right) \right.  
\nonumber\\
   && \hskip -40pt + \left. f_0(\xi_{{\bm k}'}/T)\left(1 - f_0(\xi_{\bm k}/T)\right) \delta\left(\epsilon_{\bm k} - \epsilon_{{\bm k}'} - \omega_{{\bm k}'-{\bm k}}\right) \right]\ .
\nonumber\\
\label{eq:2.5b}
\eea
\ese
Here $\xi_{\bm k} = \epsilon_{\bm k} - \mu$, and $\epsilonF = \mu(T=0)$ and $\kF = \sqrt{2m\epsilonF}$ are the Fermi energy
and Fermi wave number, respectively. We use units such that $k_{\text{B}} = \hbar = 1$.
$T_1$ is the bosonic frequency scale, which is on the order of the Debye frequency or its magnetic analogs. $n_0(x) = 1/(e^x-1)$
is the equilibrium Bose-Einstein distribution. We assume that the phonons remain in equilibrium, leaving an exact treatment of the full
coupled non-equilibrium problem for future work. Here, and throughout the paper,
$g_0$ denotes a dimensionless coupling constant of $O(1)$ that depends on the boson being exchanged. 
$\omega_{\bm k}$ is the wave-vector dependent boson frequency. For simplicity, in the main text we will take the frequency -- wave-number relation 
to be linear, as is appropriate for phonons and antiferromagnons,
\be
\omega_{\bm k} = c \vert{\bm k}\vert\quad (\text{phonons, AFMs})\ . 
\label{eq:2.6}
\ee
Scattering by helimagnons, which have a more complicated anisotropic ${\bm k}$-dependence of the frequency, see 
Ref.~\onlinecite{Belitz_Kirkpatrick_Rosch_2006a} and Appendix \ref{app:C}, can be treated in complete analogy
at the expense of more complicated integrals. In the phonon case, $c$ is the speed of sound. In the antiferromagnetic case,
it is the spin-wave stiffness coefficient. For antiferromagnets, we consider the scattering by Goldstone modes only. Overdamped
paramagnon excitations contribute a $T^2$ to the electrical resistivity that dominates over the latter at low $T$.\cite{Ueda_1977}

We note that the kernel $W$ is symmetric, and hence the collision operator $\mathfrak C$ is
self-adjoint in the space of square-integrable functions. Furthermore, ${\mathfrak C}$ has a zero eigenvalue with the corresponding
eigenfunction the constant function $\Phi({\bm k}) = \text{const.}$, which reflects electron number conservation.

For later reference we define the nearly-free-electron density of states 
\be
\NF = \kF m/2\pi^2 
\label{eq:2.7}
\ee
which will serve as a normalization factor. 

\subsubsection{Integral equations for the transport coefficients}
\label{subsubsec:II.B.2}


In antiferromagnets and helimagnets the conduction band is split by the exchange interaction,
and the function $\Phi$ depends on the sub-band index. However, 
scattering processes that couple different sub-bands are exponentially suppressed at low temperatures.
As a result, the leading contribution to the transport coefficients at low temperature comes from scattering 
within a given sub-band (`intraband scattering'), and the contributions from the two sub-bands have the 
same temperature dependence. We therefore ignore the band splitting, which just has a minor quantitative 
effect on the prefactor of the leading power law, and treat the scattering by phonons, antiferromagnons, 
and helimagnons together. Ferromagnons, which couple only electrons in different sub-bands 
(`interband scattering') will be considered in Paper II.

For the purpose of calculating the electrical and thermal conductivity, respectively, it is convenient to write
\bse
\label{eqs:2.8}
\be
\Phi({\bm k}) = \frac{e}{m}\,{\bm E}\cdot{\bm k}\,\varphi_0(\epsilon_{\bm k})\ ,
\label{eq:2.8a}
\ee
and
\be
\Phi({\bm k}) = \frac{-1}{m}\,{\bm\nabla}T\cdot{\bm k}\,\xi_{\bm k}\,\varphi_1(\epsilon_{\bm k})\ ,
\label{eq:2.8b}
\ee
\ese
and transform the linearized Boltzmann equation (\ref{eq:2.4}) into integral equations for $\varphi_0$
and $\varphi_1$. This procedure, which uses Eqs.~(\ref{eqs:2.8}) in (\ref{eq:2.5a}), performs the angular integrations,
and ignores terms that do not contribute to the low-temperature behavior, involves a substantial amount of
algebra and can be found in, e.g., Ref.~\onlinecite{Wilson_1954}. Alternatively, the same result can be
obtained by evaluating the Kubo formula in a conserving approximation.\cite{Mahan_2000} The resulting
integral equations can be written
\bse
\label{eqs:2.9}
\bea
\Lambda(\epsilon)\, \varphi_0(\epsilon) &=& -1\ , \qquad
\label{eq:2.9a}\\
\Lambda(\epsilon)\, \varphi_1(\epsilon) &=& -\epsilon\ ,
\label{eq:2.9b}
\eea
\ese
with the collision operator $\Lambda$ defined as
\be
\Lambda(\epsilon) = \int du \left[K(\epsilon,u) R_{\epsilon\to u} - K_0(\epsilon,u)\right]\ .
\label{eq:2.10}
\ee
Here, and throughout the paper, $\int du$ indicates a definite integral over all real $u$.
The defining property of the operator $R$ is
\be
R_{\epsilon\to u}\,f(\epsilon) = f(u)
\label{eq:2.11}
\ee
for any function $f$, and the kernel $K$ has three contributions,
\be
K(\epsilon,u) = K_0(\epsilon,u) - K_1(\epsilon,u) - K_2(\epsilon,u)
\label{eq:2.12}
\ee
that are defined in terms of the effective scattering potential. For both phonons and antiferromagnons
one finds
\bse
\label{eqs:2.13}
\be
K_0(\epsilon,u) = \left[n_0\left(\frac{u-\epsilon}{T}\right) + f_0\left(\frac{u}{T}\right)\right]\,V_0(u-\epsilon) 
\label{eq:2.13a}
\ee
where
\be
V_0(u) =g_0 (u/T_1)^2 \sgn(u)\,\Theta(T_1 - \vert u\vert)
\label{eq:2.13b}
\ee
\ese
with $g_0$ the dimensionless coupling constant of $O(1)$ mentioned after Eq.~(\ref{eq:1.1b}). One way to see the origin of this structure is
explained in Appendix~\ref{app:A}. For the purpose of calculating transport coefficients in the low-temperature regime, $T\ll T_1$, this becomes
\bse
\label{eqs:2.14}
\be
K_0(\epsilon,u) = \left(\frac{T}{T_1}\right)^2 k_0\left(\frac{\epsilon}{T},\frac{u-\epsilon}{T}\right)
\label{eq:2.14a}
\ee
where
\be
k_0(x,y) = g_0 \left[n_0(y) + f_0(y+x)\right] y^2 \sgn(y)\ .
\label{eq:2.14b}
\ee
The other two kernels are given by
\bea
K_1(\epsilon,u) &=& \frac{-1}{2\epsilonF}\,(u-\epsilon)\,K_0(\epsilon,u)\ ,
\label{eq:2.14c}\\
K_2(\epsilon,u) &=& \frac{1}{2}\,\left(\frac{u-\epsilon}{T_1}\right)^2 K_0(\epsilon,u)\ .
\label{eq:2.14d}
\eea
\ese
The integral equations (\ref{eqs:2.9}) are now completely defined. The transport coefficients $\sigma$,
$\sigma_h$, and $S$ are determined by their solutions via
\bse
\label{eqs:2.15}
\bea
\sigma &=& \frac{n e^2}{mT} \int d\epsilon\,w(\epsilon)\,\varphi_0(\epsilon)\ , \hskip 30pt
\label{eq:2.15a}\\
-ST\sigma/e &=& \frac{n}{mT} \int d\epsilon\,w(\epsilon)\,\varphi_1(\epsilon)\ ,
\label{eq:2.15b}\\
T\sigma_h &=& \frac{n}{mT}\int d\epsilon\,w(\epsilon)\,\epsilon\,\varphi_1(\epsilon)\ .
\label{eq:2.15c}
\eea
\ese
Here 
\bse
\label{eqs:2.16}
\bea
w(\epsilon) = -T\,\frac{\partial f_0(\epsilon/T)}{\partial\epsilon} &=& f_0(\epsilon/T) \left[1 - f_0(\epsilon/T)\right] 
\nonumber\\
&=& \frac{1}{4\cosh^2(\epsilon/2T)}
\label{eq:2.16a}
\eea
is a weight function with normalization
\be
\int d\epsilon\,w(\epsilon) = T\ .
\label{eq:2.16b}
\ee
\ese
The expressions (\ref{eqs:2.14}) through (\ref{eqs:2.15}) are valid for determining the
leading low-temperature dependence of the transport coefficients only, as their derivations neglect terms
of order $\epsilon/\epsilonF$ that translate into $T/\epsilonF$ via the temperature scaling of $\epsilon$.\cite{thermopower_footnote}

For later reference we note that the kernels obey the symmetry relations
\bse
\label{eqs:2.17}
\bea
w(\epsilon)\,K_{0,2}(\epsilon,u) &=& w(u)\,K_{0,2}(u,\epsilon)\ ,
\label{eq:2.17a}\\
w(\epsilon)\,K_1(\epsilon,u) &=& -w(u)\,K_1(u,\epsilon)\ ,
\label{eq:2.17b}
\eea
\ese
which can be checked by an explicit calculation. We further note that the structure of the kernel $K$
gives rise to three parts of the collision operator,
\bse
\label{eqs:2.18}
\be
\Lambda(\epsilon) = \Lambda_0(\epsilon) - \Lambda_1(\epsilon) - \Lambda_2(\epsilon)
\label{eq:2.18a}
\ee
with
\bea
\Lambda_0(\epsilon) &=& \int du \left[K_0(\epsilon,u) R_{\epsilon\to u} - K_0(\epsilon,u)\right]\ , \qquad
\label{eq:2.18b}\\
\Lambda_1(\epsilon) &=& \int du\,K_1(\epsilon,u) R_{\epsilon\to u}\ ,
\label{eq:2.18c}\\
\Lambda_2(\epsilon) &=& \int du\,K_2(\epsilon,u) R_{\epsilon\to u}\ .
\label{eq:2.18d}
\eea
\ese

It is useful to define functions
\bse
\label{eqs:2.19}
\be
\Gamma_n(\epsilon) = \int du\,K_n(\epsilon,u)\quad (n=0,1,2)\ .
\label{eq:2.19a}
\ee
Dimensionally, the $\Gamma_n$ are energies, and physically, relaxation rates. $\Gamma_0$ and $\Gamma_2$
are even functions of their arguments, while $\Gamma_1$ is odd. They have the scaling forms
\bea
\Gamma_0(\epsilon) &=& \frac{T^3}{T_1^2}\,\gamma_0(\epsilon/T)\ ,
\label{eq:2.19b}\\
\Gamma_1(\epsilon) &=& \frac{-T^4}{2\, T_1^2\,\epsilonF}\,\gamma_1(\epsilon/T)\ ,
\label{eq:2.19c}\\
\Gamma_2(\epsilon) &=& \frac{T^5}{2\,T_1^4}\,\gamma_2(\epsilon/T)\ ,
\label{eq:2.19d}
\eea
where
\be
\gamma_n(x) = \int dy\,y^n\,k_0(x,y)\ .
\label{eq:2.19e}
\ee
\ese
For a discussion and explicit determination of these scaling functions, see Appendix~\ref{app:B}.

\subsubsection{Properties of the collision operator}
\label{subsubsec:II.B.3}

We define a scalar product in a space of real-valued functions by means of the weight function 
$w$, Eq.~(\ref{eq:2.16a}):
\bse
\label{eqs:2.20}
\be
\langle\psi\vert\varphi\rangle = \int d\epsilon\,w(\epsilon)\,\psi(\epsilon)\,\varphi(\epsilon)\ .
\label{eq:2.20a}
\ee
In particular, if we denote by $\vert 1\rangle$ the constant function identically equal to $1$,
then the normalization of the weight function, Eqs.~(\ref{eqs:2.16}), takes the form
\be
\langle 1\vert 1\rangle = T\ .
\label{eq:2.20b}
\ee
\ese
By using the symmetry relations (\ref{eqs:2.17}) we see that $K_0$ and $K_2$ are self-adjoint
with respect to the scalar product $\langle\ \vert\ \rangle$, whereas $K_1$ is skew-adjoint. 
Now consider the eigenproblem for the part $\Lambda_0$ of the collision operator,
\be
\Lambda_0\vert\psi\rangle = \lambda\vert\psi\rangle\ .
\label{eq:2.21}
\ee
From Eq.~(\ref{eq:2.18b}) it is obvious that one of the eigenvalues is zero, $\lambda_0 = 0$,
with a constant function $\psi_0$  that is nonzero, but otherwise arbitrary, as the eigenfunction:
$\vert\psi_0\rangle \propto \vert 1\rangle$. Furthermore, this zero eigenvalue is non-degenerate, and all other eigenvalues 
are negative. To see this, we rewrite the eigenproblem as
\bea
\Lambda_0(\epsilon)\psi(\epsilon) &=& \int du\,K_0(\epsilon,u+\epsilon) \left[\psi(u+\epsilon) - \psi(\epsilon)\right]
\nonumber\\
&=& \lambda\,\psi(\epsilon)
\label{eq:2.22}
\eea
If we multiply this equation by $w(\epsilon)\psi(\epsilon)$ and integrate, we obtain
\bea
\lambda \int d\epsilon\,w(\epsilon)\,\psi(\epsilon)^2 &=&
\nonumber\\
&& \hskip -100pt =\int d\epsilon\,du\,w(\epsilon)\,K_0(\epsilon,u+\epsilon)\,\psi(\epsilon)\left[\psi(u+\epsilon) - \psi(\epsilon)\right]
\nonumber\\
&& \hskip -100pt =\int d\epsilon\,du\,w(u+\epsilon)\,K_0(u+\epsilon,\epsilon)\,\psi(\epsilon)\left[\psi(u+\epsilon) - \psi(\epsilon)\right]
\nonumber\\
&& \hskip -100pt = - \int d\epsilon\,du\,w(\epsilon)\,K_0(\epsilon,u+\epsilon)\,\psi(u+\epsilon)\left[\psi(u+\epsilon) - \psi(\epsilon)\right]
\nonumber\\
&& \hskip -100pt = \frac{-1}{2} \int d\epsilon\,du\,w(\epsilon)\,K_0(\epsilon,u+\epsilon)\left[\psi(u+\epsilon) - \psi(\epsilon)\right]^2 .
\label{eq:2.23}
\eea
Here we have used Eq.~(\ref{eq:2.17a}) to go from the second line to the third line, have relabeled the integration variables
$\epsilon \to \epsilon - u$ and $u \to -u$ to go from the third line to the fourth line, and have added the third and fourth lines to 
arrive at the last line. From the definition of $K_0$, Eqs.~(\ref{eqs:2.13}), it is easy to see that $K_0(\epsilon,u) \geq 0$, and
 $K_0(\epsilon,u) = 0$ if and only if $\epsilon = u$. Therefore, Eq.~(\ref{eq:2.23}) implies that $\lambda = 0$ if and only if
 $\psi(u+\epsilon)  = \psi(\epsilon)$ for all values of $u$ and $\epsilon$, which in turn implies that $\psi(\epsilon) \equiv \text{const}$.
 We conclude that there is only one linearly independent eigenfunction for the zero eigenvalue, and all other eigenvalues are
 negative. 
 
We will refer to the nonzero constant eigenfunction as the `zero eigenfunction', as opposed to the null function that is identically
equal to zero. Note that this zero eigenvalue of $\Lambda_0$ is different
from the zero eigenvalue of the operator ${\mathfrak C}$ in Sec.~\ref{subsubsec:II.B.1} above, which reflected electron
number conservation. The zero eigenvalue of $\Lambda_0$, by contrast,  reflects the fact that the electron momentum
is asymptotically conserved as $T\to 0$ since the excitations (phonons or magnons) that invalidate electron momentum
conservation get frozen out. This will be important in the next section, where we will see that the contributions $\Lambda_1$
and $\Lambda_2$ to the collision operator perturb the zero eigenvalue and lead to a nonzero smallest eigenvalue of the
full collision operator $\Lambda$ that vanishes as $T\to 0$. We also note that the same reasoning as above implies that
the collision operator $\mathfrak{C}$ in Eqs.~(\ref{eqs:2.5}) has a negative semi-definite spectrum. This ensures that
the perturbed zero eigenvalue of $\Lambda$ is negative.

\section{Solutions of the integral equations}
\label{sec:III}

In this section we present formally exact solutions of the integral equations (\ref{eqs:2.9}). 
Technically, what makes an exact solution possible is the fact that the
leading, in a temperature-expansion sense, part of the collision operator is the operator $\Lambda_0$ in Eqs.~(\ref{eq:2.18a})
and (\ref{eq:2.18b}), which has a zero eigenvalue with the corresponding zero eigenfunction being the constant function,
as we have seen in Sec.~\ref{subsubsec:II.B.3}. The less leading parts of the collision operator perturb the zero eigenvalue and lead to a
nonzero negative smallest eigenvalue of the full collision operator $\Lambda$ for $T>0$. This perturbed zero eigenvalue is the dominant 
contribution to the electrical conductivity and the thermopower, i.e., the two transport coefficients that are related, roughly speaking, to the 
inverse collision operator. The situation with respect to the thermal conductivity is somewhat more complicated, as we will see.

In Sec.~\ref{subsec:III.A} we construct a solution based on a spectral representation of the collision operator
$\Lambda$. The assumptions implicit in this exact solution can be eliminated, and the procedure made rigorous, if
at the expense of substantially increased mathematical complexity, see Ref.~\onlinecite{Amarel_Belitz_Kirkpatrick_2020}.
In Sec.~\ref{subsec:III.B} we give an alternative solution method using projection operators\cite{Zwanzig_1961, Mori_1965}
that yields the same result. 

Using the scalar product notation from Eqs.~(\ref{eqs:2.20}) we write the
integral equations (\ref{eqs:2.9}) in the form
\bse
\label{eqs:3.1}
\bea
\Lambda \vert\varphi_0\rangle &=& -\vert 1\rangle\ ,
\label{eq:3.1a}\\
\Lambda \vert\varphi_1\rangle &=& -\vert \epsilon\rangle\ ,
\label{eq:3.1b}
\eea
\ese
where $\vert\epsilon\rangle$ represents the linear function $f(\epsilon) = \epsilon$. 
The transport coefficients from Eqs.~(\ref{eqs:2.15}) can now be written
\bse
\label{eqs:3.2}
\bea
\sigma &=& \frac{n e^2}{mT}\,\langle\varphi_0\vert 1\rangle\ ,
\label{eq:3.2a}\\
-S T\sigma/e &=& \frac{n}{mT}\,\langle\varphi_1\vert 1\rangle\ ,
\label{eq:3.2b}\\
T\sigma_h &=& \frac{n}{mT}\,\langle\varphi_1\vert\epsilon\rangle\ .
\label{eq:3.2c}
\eea
\ese

\subsection{Solution of the integral equations I: Spectral method}
\label{subsec:III.A}
%

We assume that the collision operator $\Lambda$ has a spectral representation
\be
\Lambda = \sum_n \mu_n\,\frac{\vert e_n\rangle\langle e_n\vert}{\langle e_n\vert e_n\rangle}
\label{eq:3.3}
\ee
with discrete eigenvalues $\mu_n$ and a complete orthogonal set of right eigenvectors $\vert e_n\rangle$
and left eigenvectors $\langle e_n\vert$.\cite{LR_footnote} We have defined the collision operator
such that its spectrum is negative definite, as is customary in kinetic theory. The unit operator is represented by
\be
\mathbbm{1} = \sum_n \frac{\vert e_n\rangle\langle e_n\vert}{\langle e_n\vert e_n\rangle}\ .
\label{eq:3.4}
\ee
The solutions of the integral equations
(\ref{eqs:2.9}) will be dominated by the eigenvalue of $\Lambda$ with the smallest absolute value, which we denote
by $\mu_0$, with a corresponding (right) eigenfunction $\vert e_0\rangle$. $\mu_0$ and $\vert e_0\rangle$
can be determined by means of a systematic low-temperature expansion as follows.

From Sec.~\ref{subsubsec:II.B.2} we know that the
kernels $K_0$, $K_1$, and $K_2$ scale with different powers of $T$, and so do the corresponding
collision operators, which dimensionally are energies. If we use $T_1$ as the basic energy scale
and measure the collision operators in units of $T_1$, then $\Lambda_0$ scales as
$\Lambda_0 \sim (T/T_1)^3$, and $\Lambda_1$ and $\Lambda_2$ scale as $T^4/T_1^3\epsilonF$
and $(T/T_1)^5$, respectively. We can therefore set up a systematic low-temperature
expansion by introducing a small parameter $\alpha \sim T/T_1$. In addition, the operator
$\Lambda_1$ leads to factors of $T_1/\epsilonF$. This ratio is not necessarily small
(it is in metals, but not in, e.g., semiconductors), and we will not assume that it is. We now write
\be
\Lambda = \alpha^3 \Lambda_0 - \alpha^4 \Lambda_1 - \alpha^5\Lambda_2\ ,
\label{eq:3.5}
\ee
and put $\alpha =1$ in the end. We expand $\mu_0$ and $\vert e_0\rangle$ in
powers of $\alpha$,
\bse
\label{eqs:3.6}
\bea
\mu_0 = \alpha^3 \mu_0^{(0)} + \alpha^4\mu_0^{(1)} + \alpha^5 \mu_0^{(2)} + O(\alpha^6)\ ,
\label{eq:3.6a}\\
\vert e_0\rangle = \vert e_0^{(0)}\rangle + \alpha  \vert e_0^{(1)}\rangle + \alpha^2  \vert e_0^{(2)}\rangle + O(\alpha^3)
\label{eq:3.6b}
\eea
\ese
and consider the eigenproblem
\be
\Lambda \vert e_0\rangle = \mu_0 \vert e_0\rangle\ .
\label{eq:3.7}
\ee

Two properties of the collision operator will be very useful for analyzing this problem, viz.
\bse
\label{eqs:3.8}
\bea
\Lambda_0 \vert 1\rangle &=& \langle 1\vert \Lambda_0  = 0\ ,
\label{eq:3.8a}\\
\Lambda_1 \vert 1\rangle &=& \vert\Gamma_1\rangle = \frac{-1}{2\epsilonF}\,\Lambda_0 \vert\epsilon\rangle\ .
\label{eq:3.8b}
\eea
\ese
Equation~(\ref{eq:3.8a}) reflects the zero eigenvalue of $\Lambda_0$
that was discussed in Sec.~\ref{subsubsec:II.B.3}. 
Equation~(\ref{eq:3.8b}) follows from the relation (\ref{eq:2.14c})
between the kernels $K_1$ and $K_0$, and $\vert\Gamma_1\rangle$ represent the function $\Gamma_1$
defined by Eq.~(\ref{eq:2.19a}).
We further note that the skew-adjointness of $\Lambda_1$ implies
\bse
\label{eqs:3.9}
\be
\langle 1\vert \Lambda_1 = \frac{1}{2\epsilonF}\,\langle\epsilon\vert \Lambda_0   = - \langle\Gamma_1\vert\ .
\label{eq:3.9a}
\ee
and hence
\be
\langle 1\vert \Lambda_1 \vert 1\rangle = 0\ .
\label{eq:3.9b}
\ee
\ese

We now expand Eq.~(\ref{eq:3.7}) in powers of $\alpha$ and compare coefficients. 
To lowest (i.e., cubic) order in $\alpha$ we have, from Eq.~(\ref{eq:2.18b}) or (\ref{eq:3.8a}),
\bse
\label{eqs:3.10}
\bea
\mu_0^{(0)} &=& 0\ ,
\label{eq:3.10a}\\
\vert e_0^{(0)}\rangle &=& \vert 1\rangle \ ,
\label{eq:3.10b}\\
\langle e_0^{(0)}\vert &=& \langle 1\vert\ .
\eea
\ese
To next-leading (i.e., quartic) order in $\alpha$ we have
\be
\Lambda_0 \vert e_0^{(1)}\rangle -  \Lambda_1 \vert 1\rangle = \mu_0^{(1)} \vert 1\rangle\ .
\label{eq:3.11}
\ee
Multiplying this equation from the left with $\langle 1\vert$, and using Eqs.~(\ref{eq:3.8a})
and (\ref{eq:3.9b}) we have
\bse
\label{eqs:3.12}
\be
\mu_0^{(1)} = 0\ .
\label{eq:3.12a}
\ee
For the eigenvector at this order we thus obtain
\bea
\vert e_0^{(1)}\rangle &=& \Lambda_0^{-1} \Lambda_1 \vert 1\rangle = \frac{-1}{2\epsilonF}\,\vert\epsilon\rangle\ ,
\label{eq:3.12b}\\
\langle e_0^{(1)}\vert &=& \langle 1\vert\Lambda_1  \Lambda_0^{-1} = \frac{1}{2\epsilonF}\,\langle\epsilon\vert\ ,
\label{eq:3.12c}
\eea
\ese
where we have used Eq.~(\ref{eq:3.8b}). Note that the inverse $\Lambda_0^{-1}$ formally exists in this
context since the function $\Lambda_1\vert 1\rangle= \vert\Gamma_1\rangle$ is orthogonal to the zero 
eigenvector of $\Lambda_0$ due to symmetry, and hence the zero eigenvalue does not contribute. 
(We will come back to this point  in Sec.~\ref{subsec:III.B}.) To fifth order in $\alpha$ we have
\be
\Lambda_0 \vert e_0^{(2)}\rangle - \Lambda_1 \vert e_0^{(1)}\rangle - \Lambda_2 \vert 1\rangle = \mu_0^{(2)} \vert 1\rangle\ .
\label{eq:3.13}
\ee
Again multiplying from the left with $\langle 1\vert$ this yields
\bea
\langle 1 \vert 1 \rangle \mu_0^{(2)} &=& -\langle 1\vert\Lambda_2\vert 1\rangle - \langle 1\vert\Lambda_1\vert e_0^{(1)}\rangle
\nonumber\\
&=& - \langle 1\vert\Gamma_2\rangle - \frac{1}{2\epsilonF}\,\langle\Gamma_1\vert\epsilon\rangle\ .
\label{eq:3.14}
\eea
Here we have used Eqs.~(\ref{eq:3.12b}) and (\ref{eq:3.9a}), and the function $\Gamma_2$ is defined by
Eq.~(\ref{eq:2.19a}). If we define an average with respect to the weight function $w$ by 
\be
\langle f\rangle_w = \frac{1}{T} \int d\epsilon\, w(\epsilon)\,f(\epsilon) = \frac{\langle 1\vert f\rangle}{\langle 1\vert 1\rangle}
\label{eq:3.15}
\ee
we now have for the lowest eigenvalue
\bse
\label{eqs:3.16}
\be
\mu_0 = -\alpha^5 \left[ \langle\Gamma_2\rangle_w + \frac{1}{2\epsilonF} \langle\epsilon\Gamma_1\rangle_w \right] + O(\alpha^7)\ .
\label{eq:3.16a}
\ee
Symmetry considerations show that there are no contributions to $\mu_0$ at even powers of $\alpha$. The corresponding
right and left eigenvectors are
\bea
\vert e_0\rangle &=& \vert 1\rangle - \frac{\alpha}{2\epsilonF}\,\vert\epsilon\rangle + O(\alpha^2)\ ,
\label{eq:3.16b}\\
\langle e_0\vert &=& \langle 1\vert +\frac{\alpha}{2\epsilonF}\,\langle\epsilon\vert + O(\alpha^2)\ .
\label{eq:3.16c}
\eea
\ese

We now have the following expression for the inverse of the collision operator:
\bea
\Lambda^{-1} &=& \sum_n \frac{1}{\mu_n}\,\frac{\vert e_n\rangle\langle e_n\vert}{\langle e_n\vert e_n\rangle}
                      \approx  \frac{1}{\mu_0}\,\frac{\vert e_0\rangle\langle e_0\vert}{\langle e_0\vert e_0\rangle}
\nonumber\\
&=& \frac{1 + O(\alpha^2)}{\alpha^5\mu_0^{(2)} \langle 1\vert 1\rangle} \left[\vert 1\rangle-  \frac{\alpha}{2\epsilonF}\,\vert\epsilon\rangle + O(\alpha^2)\right]
\nonumber\\
&& \hskip 45pt \times  \left[\langle 1\vert +  \frac{\alpha}{2\epsilonF}\,\langle\epsilon\vert + O(\alpha^2)\right]\ .\qquad
\label{eq:3.17}                      
\eea    
For $\varphi_0$, i.e., the solution of Eq.~(\ref{eq:3.1b}), this yields the scaling behavior
\bse
\label{eqs:3.18}
\be
\vert\varphi_0\rangle \sim \frac{1+O(\alpha^2)}{\alpha^5}\,\vert 1\rangle + \frac{1}{\alpha^4}\,\vert\epsilon\rangle + O(1/\alpha^3)\ .
\label{eq:3.18a}
\ee
Note that the $\alpha$-expansion is singular with respect to the basic $\alpha^3$-scaling of $\Lambda_0$. This is a consequence 
of the zero eigenvalue of $\Lambda_0$. $\Lambda_1$ and $\Lambda_2$ perturb the zero eigenvalue and lead to a finite inverse 
$\Lambda^{-1}$, but since they are of higher order in $\alpha$ that inverse diverges as $\alpha\to 0$. Explicitly we have for the 
leading low-temperature behavior of $\varphi_0$
\be
\vert\varphi_0\rangle_{T\to 0} = \frac{1}{\alpha^5}\,\phi_0 \vert 1\rangle -\frac{1}{\alpha^4}\,\phi_0\,\frac{1}{2\epsilonF}\,\vert\epsilon\rangle + O(1/\alpha^3)\ ,
\label{eq:3.18b}
\ee                   
where
\be
\phi_0 = \frac{-1}{\mu_0^{(2)}} = \frac{1}{\langle\Gamma_2\rangle_w + \langle\epsilon\Gamma_1\rangle_w/2\epsilonF} \ .
\label{eq:3.18c}
\ee
The integrals that determine the average rates in Eq.~(\ref{eq:3.18c}) can be performed, see Appendix~\ref{app:B}. We find
\bea
\langle\Gamma_2\rangle_w &=& 120\, \zeta(5)\, T^5/T_1^4\ ,
\label{eq:3.18d}\\
\langle\epsilon\Gamma_1\rangle_w &=& 60\, \zeta(5)\, T^5/\epsilonF T_1^2\ ,
\label{eq:3.18e}
\eea
with $\zeta$ the Riemann zeta function. If we finally put $\alpha=1$, this yields 
\bea
\phi_0 &=& \frac{1}{120\, \zeta(5) g_0}\,\frac{1}{1 + T_1^2/4\epsilonF^2}\,\frac{T_1^4}{T^5}\ ,
\label{eq:3.18f}\\
\vert\varphi_0\rangle &=& \phi_0 \vert 1\rangle - \phi_0\,\frac{1}{2\epsilonF}\,\vert\epsilon\rangle + O(1/T^3)
\label{eq:3.18g}
\eea
\ese
where the powers of $T/T_1$ correspond to the powers of $\alpha$  in Eq.~(\ref{eq:3.18a})
as they must. 

We next discuss the function $\varphi_1$ that is the solution of Eq.~(\ref{eq:3.1b}). The inhomogeneity
$\vert\epsilon\rangle$ scales as $T$ and hence carries a factor of $\alpha$ in our counting scheme, so we rewrite the
integral equation as
\be
\Lambda\vert\varphi_1\rangle = -\alpha\vert\epsilon\rangle\ .
\tag{3.1b'}
\ee
The solution has a `hydrodynamic' contribution\cite{hydrodynamic_footnote} that is related to the perturbed zero eigenvalue and is obtained by
operating with $\Lambda^{-1}$ as given in Eq.~(\ref{eq:3.17}) on $\vert\epsilon\rangle$:
\be
\vert\varphi_1\rangle^{\text{hyd}} = \frac{1}{\alpha^3}\,\phi_0\,\frac{\langle\epsilon^2\rangle_w}{2\epsilonF} \left[\vert 1\rangle - \frac{\alpha}{2\epsilonF}\vert\epsilon\rangle + O(\alpha^2)\right]
\label{eq:3.19}
\ee
In addition, there is a `non-hydrodynamic' or `kinetic' contribution that is unrelated to the perturbed zero eigenvalue.
To lowest order in $\alpha$ it is given by the solution of
\be
\Lambda_0 \vert\varphi_1\rangle^{\text{kin}} = \frac{-1}{\alpha^2}\,\vert\epsilon\rangle\ ,
\label{eq:3.20}
\ee
which exists since the inhomogeneity in Eq.~(\ref{eq:3.20}) is orthogonal to the zero eigenvector $\vert 1\rangle$. 
Equations (\ref{eq:2.18b}) and (\ref{eqs:2.14}) imply that the vector
\bse
\label{eqs:3.21}
\be
\vert\varphi_1\rangle^{\text{kin}} = \frac{1}{\alpha^2}\, \vert h\rangle
\label{eq:3.21a}
\ee
represents a function $h(\epsilon)$ that has the scaling form
\be
h(\epsilon) = \left(\frac{T_1}{T}\right)^2 {\mathfrak h}(\epsilon/T)
\label{eq:3.21b}
\ee
with $\mathfrak h$ the solution of
\be
\int dy\,k_0(x,y-x)\,{\mathfrak h}(y) - \gamma_0(x)\,{\mathfrak h}(x) = -x\ .
\label{eq:3.21c}
\ee
\ese
Here $k_0$ and $\gamma_0$ are given by Eqs.~(\ref{eq:2.14b}) and (\ref{eq:2.19c}), respectively. 
$\vert\varphi_1\rangle$, up to $O(1/\alpha^2)$, thus has the form
\be
\vert\varphi_1\rangle =  \frac{1}{\alpha^3}\,\phi_0\,\frac{\langle\epsilon^2\rangle_w}{2\epsilonF} \vert 1\rangle
     -  \frac{1}{\alpha^2}\,\phi_0\,\frac{\langle\epsilon^2\rangle_w}{4\epsilonF^2} \vert \epsilon\rangle + \frac{1}{\alpha^2}\vert h\rangle\ .
\label{eq:3.22}
\ee
The non-hydrodynamic or kinetic part $h$, which is defined by Eqs.~(\ref{eqs:3.21}), we have been unable to express explicitly, but its existence
and scaling behavior are guaranteed. Note that the kinetic part contributes only to the contribution that is orthogonal to the
zero eigenvector $\vert 1\rangle$. It scales as $1/\alpha^2$, as does the hydrodynamic contribution to that part of $\vert\varphi_1\rangle$.
However, the latter also carries a factor of $(T_1/\epsilonF)^2$. In systems where $T_1/\epsilonF \ll 1$ the non-hydrodynamic contribution 
thus dominates the hydrodynamic one.

\subsection{Solution of the integral equations II: Projector method}
\label{subsec:III.B}
%

We now give an alternative method for deriving Eqs.~(\ref{eqs:3.18}) and (\ref{eqs:3.21}). We define an operator $P$
that projects onto the one-dimensional subspace spanned by the zero eigenfunction $\vert 1\rangle$ of $\Lambda_0$,
\bse
\label{eqs:3.23}
\be
P = \frac{1}{\langle 1\vert 1 \rangle}\,\vert 1\rangle\langle 1\vert\ ,
\label{eq:3.23a}
\ee
and a second projector $P_{\perp}$ that projects onto the complement of that subspace,
\be
P_{\perp} = \mathbbm{1} - P\ .
\label{eq:3.23b}
\ee
\ese
In addition to the projector property 
\bse
\label{eqs:3.24}
\be
P^2 = P \quad,\quad P_{\perp}^2 = P_{\perp}\ ,
\label{eq:3.24a}
\ee
the projectors have the following properties:
\bea
P\Lambda_0 &=& \Lambda_0 P = 0\ ,
\label{eq:3.24b}\\
P_{\perp} \Lambda_1 \vert 1\rangle &=& \Lambda_1 \vert 1\rangle = \vert\Gamma_1\rangle\ ,
\label{eq:3.24c}\\
P\Lambda_1 P &=& 0\ ,
\label{eq:3.24d}\\
P \Lambda_2 P &=& \langle\Gamma_2\rangle_w P\ .
\label{eq:3.24e}
\eea
\ese
Equation (\ref{eq:3.24b}) reflects the zero eigenvalue of $\Lambda_0$, and Eqs.~(\ref{eq:3.24c}, \ref{eq:3.24d})
reflect the skew-adjointness of $\Lambda_1$. We further define
\be
\Lambda_{0\perp} = P_{\perp} \Lambda_0 P_{\perp}
\label{eq:3.25}
\ee
which possesses an inverse since $P_{\perp}$ removes the zero eigenvalue of $\Lambda_0$. 
As we will see, within the framework of the projector method the inverse of the collision operator
$\Lambda_0$ appears only in the form of $\Lambda_{0\perp}^{-1}$, which manifestly exists. 

Now we write the integral equation (\ref{eq:2.9a}) for $\varphi_0$ in the form
\bea
-\vert 1\rangle &=& \Lambda (P + P_{\perp}) \vert\varphi_0\rangle
\nonumber\\
&=& \alpha^3\left[\Lambda_0 P \vert\varphi_0\rangle + \Lambda_0 P_{\perp}  \vert\varphi_0\rangle - \alpha \Lambda_1 P  \vert\varphi_0\rangle \right.
\nonumber\\
&& \left.  - \alpha \Lambda_1 P_{\perp}  \vert\varphi_0\rangle  - \alpha^2 \Lambda_2 P  \vert\varphi_0\rangle - \alpha^2 \Lambda_2 P_{\perp}  \vert\varphi_0\rangle \right]\ .
\nonumber\\
\label{eq:3.26}
\eea
Operating on this identity from the left with $P$ and using Eqs.~(\ref{eqs:3.24}) yields 
\bse
\label{eqs:3.27}
\bea
-\vert 1\rangle &=& -\alpha^3\left\{\left[\alpha P \Lambda_1 P_{\perp} + O(\alpha^2)\right] P_{\perp} \vert\varphi_0\rangle \right.
\nonumber\\
&& \hskip 80pt \left. - \alpha^2 \langle\Gamma_2\rangle_w P \vert\varphi_0\rangle\right\}\ . \qquad
\label{eq:3.27a}
\eea
Similarly, operating on Eq.~(\ref{eq:3.26}) with $P_{\perp}$ yields
\be
0 = \left[\Lambda_{0\perp} + O(\alpha)\right] P_{\perp} \vert\varphi_0\rangle - \left[\alpha P_{\perp} \Lambda_1 + O(\alpha^2)\right] P\vert\varphi_0\rangle\ .
\label{eq:3.27b}
\ee
\ese
Using Eq.~(\ref{eq:3.27b}) in Eq.~(\ref{eq:3.27a}) to express $P_{\perp}\vert\varphi_0\rangle$ in terms of $P\vert\varphi_0\rangle$ we obtain
\be
-\vert 1\rangle = -\alpha^5 \left(P \Lambda_1 P_{\perp} \Lambda_{0\perp}^{-1} P_{\perp} \Lambda_1 + \langle\Gamma_2\rangle_w \right) P\vert\varphi_0\rangle + O(\alpha^3)\ .
\label{eq:3.28}
\ee
But $P\vert\varphi_0\rangle \propto \vert 1\rangle$, so we write 
\bse
\label{eqs:3.29}
\be
P\vert\varphi_0\rangle = \frac{1}{\alpha^5}\,\phi_0 \vert 1\rangle + O(1/\alpha^3)\ .
\label{eq:3.29a}
\ee 
Multiplying Eq.~(\ref{eq:3.28}) from the left with $\langle 1\vert$ and using Eqs.~(\ref{eqs:3.24}) as well as Eqs.~(\ref{eq:3.8a}) and (\ref{eq:3.9a})
we obtain $\phi_0$ as written in Eq.~(\ref{eq:3.18c}). Inserting this in Eq.~(\ref{eq:3.27b}) and solving for $P_{\perp}\vert\varphi_0\rangle$
yields, with the help of Eq.~(\ref{eq:3.8b}), 
\be
P_{\perp}\vert\varphi_0\rangle = \frac{-1}{\alpha^4}\,\phi_0\,\frac{1}{2\epsilonF}\,\vert\epsilon\rangle + O(1/\alpha^3)\ .
\label{eq:3.29b}
\ee
\ese
Since $\vert\varphi_0\rangle = P\vert\varphi_0\rangle + P_{\perp}\vert\varphi_0\rangle$ we now have Eq.~(\ref{eq:3.18b}) for $\vert\varphi_0\rangle$.

We now use the same method for determining $\varphi_1$. We write Eq.~(\ref{eq:2.9b}) in the form
\bea
-\alpha\vert\epsilon\rangle &=& \Lambda (P + P_{\perp}) \vert\varphi_1\rangle
\nonumber\\
&=& \alpha^3\left[ \Lambda_0 P \vert\varphi_1\rangle + \Lambda_0 P_{\perp}  \vert\varphi_1\rangle - \alpha \Lambda_1 P  \vert\varphi_1\rangle
    \right.
\nonumber\\
&& \left.   - \alpha \Lambda_1 P_{\perp}  \vert\varphi_1\rangle  - \alpha^2 \Lambda_2 P  \vert\varphi_1\rangle - \alpha^2 \Lambda_2 P_{\perp}  \vert\varphi_1\rangle \right]\ .
\nonumber\\
\label{eq:3.30}
\eea
Operating from the left with $P$ and using Eqs.~(\ref{eqs:3.24}) we obtain
\bse
\label{eqs:3.31}
\be
0 = -\left[\alpha P \Lambda_1 P_{\perp} + O(\alpha^2)\right] P_{\perp}\vert\varphi_1\rangle - \alpha^2 \langle\Gamma_2\rangle_w P\vert\varphi_1\rangle\ .
\label{eq:3.31a}
\ee
Similarly operating on Eq.~(\ref{eq:3.30}) with $P_{\perp}$ and solving for $P_{\perp}\vert\varphi_1\rangle$ we find
\be
P_{\perp}\vert\varphi_1\rangle = -\Lambda_{0\perp}^{-1}\vert\epsilon\rangle + \Lambda_{0\perp}^{-1}\left[\alpha P_{\perp} \Lambda_1 + O(\alpha^2)\right] P\vert\varphi_1\rangle\ .
\label{eq:3.31b}
\ee
\ese
Using Eq.~(\ref{eq:3.31b}) in (\ref{eq:3.31a}), and using Eqs.~(\ref{eqs:3.24}) again, we find
\bea
P\vert\varphi_1\rangle &=& \frac{1}{\alpha^3}\,\frac{1}{ 2\epsilonF \langle\Gamma_2\rangle_w}\,\frac{1}{ \langle 1\vert 1\rangle}\,
\nonumber\\
&&\hskip 0pt \times \left[\langle\epsilon\vert\epsilon\rangle
   - \alpha \langle\epsilon\vert \Lambda_1 P\vert\varphi_1\rangle + O(\alpha^2)\right] \vert 1\rangle\ .\qquad
\label{eq:3.32}
\eea 
Multiplying this from the left with $\langle\epsilon\vert \Lambda_1$ we can solve for the matrix element $\langle\epsilon\vert \Lambda_1 P\vert\varphi_1\rangle$ to find
\be
\langle\epsilon\vert \Lambda_1 P\vert\varphi_1\rangle = \frac{1}{\alpha^3}\,\frac{\phi_0}{2\epsilonF}\, \langle\epsilon\vert\epsilon\rangle \langle\epsilon\Gamma_1\rangle_w\ .
\label{eq:3.33}
\ee
Equation~(\ref{eq:3.32}) now yields
\bse
\label{eqs:3.34}
\be
P\vert\varphi_1\rangle = \frac{1}{\alpha^3}\,\phi_0\frac{\langle\epsilon^2\rangle_w}{2\epsilonF} \,\vert 1\rangle + O(1/\alpha)\ .
\label{eq:3.34a}
\ee
Using this in Eq.~(\ref{eq:3.31b}) we find
\be
P_{\perp}\vert\varphi_1\rangle =  - \frac{1}{\alpha^2}\,\phi_0  \,\frac{\langle\epsilon^2\rangle_w}{4\epsilonF^2}\,\vert\epsilon\rangle
     -  \frac{1}{\alpha^2}\,\vert h\rangle + O(1/\alpha)
\label{eq:3.34b}
\ee
\ese     
with $\vert h\rangle$ from Eqs.~(\ref{eqs:3.21}).
Adding Eqs.~(\ref{eq:3.34a}) and (\ref{eq:3.34b}) we recover Eq.~(\ref{eq:3.22}) for $\varphi_1$. 

\subsection{The transport coefficients}
\label{subsec:III.C}

From Eqs.~(\ref{eqs:3.18}) and (\ref{eq:3.2a}) we see that the electrical conductivity scales as
$\sigma \sim 1/\alpha^5 \sim T_1^4/T^5$. Explicitly, we have
\be
\sigma = \frac{n e^2}{m}\,\phi_0\, \frac{1}{\alpha^5}\, \left[1 + O(\alpha^2)\right]\ .
\label{eq:3.35}
\ee
Here $\phi_0$, which plays the role of the transport relaxation time relevant for the electrical conductivity, 
is given by Eq.~(\ref{eq:3.18c}). Putting $\alpha=1$, we thus have for the leading low-temperature contribution
\bse
\label{eqs:3.36}
\be
\sigma(T\to 0) = \frac{n e^2}{m}\,\frac{1}{120\,\zeta(5) g_0}\,\frac{1}{1 + T_1^2/4\epsilonF^2}\,\frac{T_1^4}{T^5} + O(1/T^3)\ .
\label{eq:3.36a}
\ee
This result is more commonly written in terms of the electrical resistivity $\rho = 1/\sigma$,
\be
\rho(T\to 0) = \frac{m}{n e^2}\,120\, \zeta(5) g_0\left(1 + T_1^2/4\epsilonF^2\right) T^5/T_1^4 + O(T^7)\ .
\label{eq:3.36b}
\ee
\ese
Note that this result is exact, including the prefactor of the Bloch $T^5$ behavior. 

The thermopower $S$ scales as $S \sim \alpha \sim T/\epsilonF$. 
By using Eq.~(\ref{eq:3.22}) in Eq.~(\ref{eq:3.2b}), $\langle\epsilon^2\rangle_w = \pi^2 T^2/3$, and putting $\alpha  = 1$, we obtain 
\be
-eS(T\to 0) = \frac{\pi^2}{6}\,\frac{T}{\epsilonF} + O(T^3)\ .
\label{eq:3.37}
\ee
This result, including the prefactor, is also exact.

Finally, the heat conductivity scales as $\sigma_h \sim 1/\alpha^2 \sim 1/T^2$.
By using Eq.~(\ref{eq:3.22}) in (\ref{eq:3.2c}) we have explicitly
\be
\sigma_h = \frac{1}{\alpha^2}\,\frac{n}{mT} \left(\frac{1}{T}\langle h\vert\epsilon\rangle - \phi_0\,\frac{\langle\epsilon^2\rangle^2}{4\epsilonF^2}\right)\ .
\label{eq:3.38}
\ee
with the function $h$ from Eqs.~(\ref{eqs:3.21}). 
The result for the heat conductivity is often written as an expression for $\sigma_h/T$, which dimensionally is an inverse
relaxation rate, as is the electrical conductivity $\sigma$.  (To the extent that the specific heat is linear in $T$, $\sigma_h/T$ is proportional
to the heat diffusivity.) We therefore write our final result, putting $\alpha=1$, as
\bse
\label{eqs:3.39}
\bea
\sigma_h(T\to 0)/T &=& \frac{n}{m}\frac{1}{g_0}\left( \eta - \frac{\pi^4/9}{120 \zeta(5)}\,
   \frac{T_1^2/4\epsilonF^2}{1+T_1^2/4\epsilonF^2}\right)\frac{T_1^2}{T^3} 
   \nonumber\\
   && \hskip 70pt + O(1/T)\ ,
\label{eq:3.39a}
\eea
where
\be
\eta =  g_0\int d\epsilon\,\frac{\epsilon\,{\mathfrak h}(2\epsilon)}{\cosh^2\epsilon}\ ,
\label{eq:3.39b}
\ee
\ese
with ${\mathfrak h}$ from Eqs.~(\ref{eqs:3.21}), is a number independent of $g_0$. 
This result for the heat conductivity, Eqs.~(\ref{eqs:3.39}), is also exact. It displays the well known $1/T^3$ behavior, 
as opposed to the Bloch $1/T^5$ scaling of the electrical conductivity.\cite{Ziman_1960, Wilson_1954}

\section{Discussion}
\label{sec:IV}

In summary, we have given exact solutions of the integral equations that determine the
electrical conductivity, the thermopower, and the thermal conductivity in simple metals
in the low-temperature limit. Our method establishes that the well-known power-law
temperature dependences of these transport coefficients are exact at asymptotically
low temperatures, and it also yields the prefactors of the power laws exactly. We
emphasize that the method is very general. It relies only on the fact that the electron
momentum is conserved at zero temperature, and still approximately conserved at low 
temperature. The origin and quality of the excitations that scatter the electrons is
irrelevant, as is demonstrated by the fact that the method works equally well for 
scattering by phonons, antiferromagnons, and helimagnons. In Paper II we will
show that it also works for ferromagnons. It furthermore is not restricted to scattering
by weakly damped particle-like excitations, it is equally applicable to scattering by
overdamped excitations such as paramagnons.

We conclude by discussing various aspects of our procedure and our results.
              
\subsection{Relation to previous treatments}
\label{subsec:IV.A}

We briefly discuss the relation between our analysis and previous treatments of the problem.

\subsubsection{Solution of the integral equation}
\label{subsubsec:IV.A.1}

One popular method for solving the linearized Boltzmann equation for the electrical conductivity, or an equivalent integral equation, is to
transform it into an algebraic equation by replacing all energy-dependent rates by constants.\cite{Mahan_2000}
As the constant solution for the function $\varphi_0$, Eq.~(\ref{eq:3.18g}), shows, this is qualitatively correct,
but obviously misses the contribution from the kernel $K_1$, since $\langle\Gamma_1\rangle = 0$. 
Another method is to construct variational solutions,\cite{Wilson_1954} and the thermal transport coefficients have
been calculated by adding quenched disorder, assuming the validity of Matthiessen's rule, and interchanging
limits.\cite{Wilson_1954} All of these approximations are uncontrolled. Whereas they do yield the correct 
low-temperature dependence, even this conclusion can be drawn only by solving the integral equations exactly, 
as we have done in this paper, and on occasion it has been doubted that they do.\cite{Mahan_2000} 
The approximations obviously provide no control over the prefactor of the power laws. As mentioned above,
the most commonly used approximation for the electrical conductivity misses the $\Gamma_1$ contribution,
i.e., the $T_1^2/4\epsilonF^2$ term in the denominator of Eq.~(\ref{eq:3.18f}), even if the rates are replaced
by their energy averages with the appropriate weight. The Wilson-Sondheimer result for the thermopower, 
$-e S(T\to 0) = \pi^2 T/3\epsilonF$,\cite{Wilson_1954, Ziman_1960} has the correct temperature dependence, 
but the prefactor is too large by a factor of $2$. The thermal conductivity is sometimes argued to be given
by the single-particle relaxation rate, since thermal relaxation does not favor backscattering events as the
electrical conductivity does.\cite{Ziman_1960} While it is true that the heat diffusivity and the single-particle
rate have the same temperature scaling, see Eqs.~(\ref{eq:3.39a}) and (\ref{eq:B.4a}), this statement is
misleading: The solution for the function $\varphi_1$, Eq.~(\ref{eq:3.22}), is not related to the single-particle
rate. 

Our technique for a formally exact solution was developed in Ref.~\onlinecite{Amarel_Belitz_Kirkpatrick_2020}. 
The same reference showed how to eliminate various assumptions (most prominently how to prove the existence and properties
of the spectral representation of the collision operator) and make the treatment mathematically rigorous.
A non-rigorous derivation of the Bloch law for the electrical conductivity based on spectral methods was 
given earlier in Ref.~\onlinecite{Belitz_Kirkpatrick_2010a}.

Our general technique of taking advantage of the perturbed zero eigenvalue of the collision operator,
whose eigenvector we refer to as the hydrodynamic mode, is based on classical kinetic theory, see, e.g., 
Ref.~\onlinecite{Dorfman_vanBeijeren_Kirkpatrick_2020}. In the quantum case the limit $T\to 0$ provides
perturbative control that allows for an exact determination of the low-temperature behavior. Both the spectral
representation method in Sec.~\ref{subsec:III.A} and the projector method in Sec.~\ref{subsec:III.B} allow
for a controlled expansion in the parameter $\alpha \sim T$, even though the underlying integral equation
cannot be solved by iteration, due to the zero eigenvalue of $\Lambda_0$. 

\subsubsection{Origin of the skew-symmetric kernel}
\label{subsubsec:IV.B.2}

The frequency dependence of the kernels that is not related to the distribution functions derives from averages of the 
spectrum $V''$ of the effective potential over energy shells at fixed distances from the chemical potential. For the 
intraband case this average has the form
\be
\frac{1}{\NF^2 V^2} \sum_{{\bm k},{\bm p}} V''({\bm k}-{\bm p},u)\,\delta(\xi_{\bm k} - \epsilon)\,\delta(\xi_{\bm p} - \epsilon - u)
\label{eq:4.1}
\ee
The distance $\epsilon$ from the Fermi surface scales as the temperature by virtue of the weight function $w$ defined in Eq.~(\ref{eq:2.16a}).
The $\epsilon$ in the arguments of the $\delta$-functions therefore leads only to effects of $O(T/\epsilonF)$ that do not
contribute to the leading low-$T$ behavior.\cite{thermopower_footnote_2} The $u$ in the argument of the second $\delta$-function
leads to the kernel $K_1$, and in addition to corrections to the limits of the wave-number integration that again do not
contribute to the leading low-$T$ behavior. The kernel $K_1$ was neglected in Ref.~\onlinecite{Belitz_Kirkpatrick_2010a}, its role
was discussed in Ref.~\onlinecite{Amarel_Belitz_Kirkpatrick_2020}.

It is important that the kernel $K_1$, and hence the contribution $\Lambda_1$ to the colllision operator, is skew-adjoint with respect 
to the scalar product defined in Eq.~(\ref{eq:2.20a}), while $\Lambda_0$ and $\Lambda_2$ are self-adjoint. By contrast, the collision
operator $\mathfrak{C}$ in Eq.~(\ref{eq:2.5a}) is self-adjoint. This is because $\mathfrak{C}$ and $\Lambda$ are defined on
different function spaces, with different scalar products.

\subsection{Technical aspects of the exact solution}
\label{subsec:IV.B}

Our formally exact treatment relies on several assumptions, most importantly, 
the existence of a spectral representation of the collision operator with a discrete spectrum, Eq.~(\ref{eq:3.3}). This is not assured, 
since the collision operator is not self-adjoint with respect to the chosen function space. 
The projector method of Sec.~\ref{subsec:III.B} relies on the same assumptions, since the very construction of the
projectors hinges on the properties of the spectrum. These assumptions can be rigorously proven, if desired,
as was shown in Ref.~\onlinecite{Amarel_Belitz_Kirkpatrick_2020}.

As was mentioned after Eq.~(\ref{eq:2.5b}), the integral equations (\ref{eqs:2.9}) that are our starting point
are derived from the Boltzmann equation under the unphysical assumption that the bosons remain in
thermal equilibrium. It is conceivable that our method can also be used to construct an exact solution of
the full coupled non-equilibrium problem. This is left for a future project.

\subsection{Remarks concerning the thermal conductivity}
\label{subsec:IV.C}

The heat conductivity as given by Eq.~(\ref{eq:2.15c}) or, more generally, by the heat current -- heat current
Kubo function, is the correlation function of the kinetic part of the energy current only. In addition, there is a
potential contribution to the energy current that is not explicit in the Kubo formula. This contribution is hidden
in the Fermi-liquid parameters that enter the Kubo expression for the heat conductivity, as Landau Fermi-liquid
theory maps the interacting problem, where the potential energy is present, onto a problem of non-interacting
quasiparticles. 

A qualitative difference between the heat conductivity on one hand, and the electrical conductivity or the thermopower
on the other, is that the leading low-temperature dependence of the heat conductivity is not determined by the
perturbed zero eigenvalue alone. As is obvious from Eq.~(\ref{eq:3.22}), the non-hydrodynamic, or kinetic, 
part of the function $\varphi_1$ contributes equally to the heat conductivity (but not to the thermopower). 
The kinetic part is given by the scaling function $\mathfrak{h}$, which in turn is the solution of the integral
equation (\ref{eq:3.21c}), which is not known explicitly. Consequently, the final result for the heat conductivity,
Eq.~(\ref{eq:3.39a}), depends on the number $\eta$ which is not known explicitly.

The scaling behavior of the kinetic contribution to the heat conductivity can be immediately deduced from
Eq.~(\ref{eq:3.20}): The collision operator $\Lambda_0$ scales as the single-particle rate $\Gamma_0$,
and hence we have, in  scaling sense, $\varphi_1^{\text{kin}}(\epsilon) \sim \epsilon/\Gamma_0(\epsilon) = \epsilon\,T_1^2/T^3\,\gamma_0(\epsilon/T)$.
For the heat conductivity Eq.~(\ref{eq:3.2c}) this yields $\sigma_h^{\text{kin}}/T \sim 1/\langle\Gamma_0\rangle_w \propto T_1^2/T^3$, in agreement
with Eq.~(\ref{eq:3.39a}). Here $\langle\Gamma_0\rangle_w$ is the averaged single-particle rate from
Eq.~(\ref{eq:B.4a}). The hydrodynamic contribution has the same temperature scaling, as the power-counting
arguments in Sec.~\ref{sec:III} show, but carries a factor of $(T_1/\epsilonF)^2$. We emphasize, however,
that these simple arguments are valid in a scaling sense only, see the remarks in Sec.~\ref{subsubsec:IV.A.1}.
               
\appendix

\section{Effective potentials}
\label{app:A}
In this appendix we list the dynamical potentials that describe the effective electron-electron interaction due to boson exchange
and explain their origins. All of the potentials have the form
\bse
\label{eqs:A.1}
\be
V({\bm k},z) = \frac{v({\bm k})}{\NF^2}\,\chi({\bm k},z)\ .
\label{eq:A.1a}
\ee
Here $\chi$ is the susceptibility of the relevant bosonic excitation as a function of a wave vector ${\bm k}$
and a complex frequency $z$. Our normalization is such that $\chi$ is dimensionally a density of states.
For undamped propagating excitations it has the form
\be
\chi({\bm k},z) \propto \frac{1}{\omega_0^2({\bm k}) - z^2}
\label{eq:A.1b}
\ee
\ese
with $\omega_0({\bm k})$ the resonance frequency of the relevant excitation. The numerator, which
dimensionally is an energy divided by a volume, also depends on the nature of the excitation and may be ${\bm k}$ dependent,
see below. $v({\bm k})$ is a dimensionless coupling factor whose ${\bm k}$-dependence depends on the
number of Fermi-surface sheets and the nature of the scattering process, as follows.

In the phonon case there is only one Fermi surface, the physically most obvious choice for $\chi$ arguably is the 
density susceptibility, and $v({\bm k} \to 0) = v_0$ with $v_0>0$ a number of $O(1)$. 
(However, this choice for $\chi$, and the resulting constant coupling factor, obscures the presence of a
gradient squared that is built into the density susceptibility, see the discussion in Sec.~\ref{app:A.1a} below.)
In the magnetic cases the relevant susceptibilities are related to the spin
density, but the scattering processes are more complicated because the exchange
interaction splits the Fermi surface, resulting in two sub-bands separated in energy space by the
exchange splitting $\lambda$. The spin fluctuations are generalized phases, and electrons
within the same sub-band can couple only to the gradient of the phase. For scattering within
the same sub-band (`intraband scattering') the coupling function therefore has the form
\bse
\label{eqs:A.2}
\be
v({\bm k}\to 0) = v_0 ({\bm k}/\kF)^2 \quad (\text{intraband scattering})\ .
\label{eq:A.2a}
\ee
Whereas, if the scattering is between sub-bands (`interband scattering'), then the two phases are
distinct, the coupling does not require a gradient, and one has
\be
v({\bm k}\to 0) = v_0 \quad (\text{interband scattering})\ .
\label{eq:A.2b}
\ee
\ese

In the magnetic cases, where the Fermi surface is split, the potential $V$, the susceptibility $\chi$,
and the coupling $v$ all are $2\times 2$ matrices in the sub-band index. 
It is physically obvious that the low-temperature dependence of the scattering rates will
be qualitatively different for interband and intraband scattering, respectively. Intraband scattering
will lead to power-law relaxation rates in the zero-temperature limit, whereas interband scattering will lead to
exponentially suppressed rates for temperatures $T\ll T_0$, with $T_0$ a temperature scale determined
by $\lambda$. Whenever interband scattering processes 
are present they will therefore dominate the low-temperature rates. The relevant matrix elements then are
the diagonal ones, and the contributions from the two 
sub-bands will simply add. This is the case for scattering by antiferromagnons or helimagnons. The
ferromagnetic case is special since the magnons do not couple electrons within the same sub-band.
As a result, one has to deal with coupled equations for the relaxation rates in the two bands, and
the low-temperature rates are exponentially small with power-law prefactors. We will discuss the
scattering problem for this case in Paper II.

We now specify the relevant susceptibilities for all four cases. 

\subsection{Susceptibilities}
\label{app:A.1}

\subsubsection{Phonons}
\label{app:A.1a}

The relevant susceptibility is the charge or number density susceptibility $\chi_n$\cite{Forster_1975, Fetter_Walecka_1971}
\bse
\label{eqs:A.3}
\be
\chi_n({\bm k},z) = \frac{n_i}{M_i c^2}\,\frac{\omega_{\text{ph}}^2({\bm k})}{\omega_{\text{ph}}^2({\bm k}) - z^2}\ ,
\label{eq:A.3a}
\ee
where $M_i$ and $n_i$ are the ionic mass and density, respectively, $c$ is the speed of sound, and
the resonance frequency
\be
\omega_{\text{ph}}({\bm k}) = c\,\vert{\bm k}\vert
\label{eq:A.3b}
\ee
\ese
displays the dispersion relation for acoustic phonons. 

It is illustrative to recall that, in a crystal at $T=0$, density fluctuations $\delta n$ are given by the divergence
of the displacement vector ${\bm u}$, $\delta n = n_i\,{\bm\nabla}\cdot{\bm u}$, and the displacement 
susceptibility $\chi_u$ is thus given by
\bse
\label{eqs:A.4}
\be
\chi_u({\bm k},z) = \frac{1/n_i M_i}{\omega_{\text{ph}}^2({\bm k}) - z^2}
\label{eq:A.4a}
\ee
The divergence of the static displacement susceptibility in the long-wavelength limit,
\be
\chi_u({\bm k},z=0) = \frac{1/n_i M_i c^2}{{\bm k}^2}
\label{eq:A.4b}
\ee
\ese
reflects the nature of the phonon as the Goldstone mode that results from the spontaneously broken
translational symmetry.

The fact that the density susceptibility determines the effective potential for electron-phonon scattering,
rather than the Goldstone susceptibility $\chi_u$, can be interpreted as follows. Physically, the displacement
is a generalized phase, which electrons cannot couple to directly. The coupling is to the gradient of the
phase, i.e., the density, which is reflected in the extra factor of ${\bm k}^2$ in the numerator of $\chi_n$ compared
to $\chi_u$. This is the reason why the coupling $v({\bm k})$ in Eq.~(\ref{eq:A.1a}) is a constant in the
electron-phonon case, even though the scattering is of intraband nature. If one used the Goldstone
susceptibility $\chi_u$, properly normalized, instead of $\chi_n$, then the gradient squared would be
provided by $v({\bm k})$, i.e., by Eq.~(\ref{eq:A.2a}) with $\kF$ replaced by the Debye wave number.

\subsubsection{Antiferromagnons}
\label{app:A.1b}

In antiferromagnets, the relevant susceptibility is the staggered magnetization susceptibility 
whose most singular part is\cite{Forster_1975, Chaikin_Lubensky_1995}
\bse
\label{eqs:A.5}
\be
\chi_{\text{AFM}}({\bm k},z) = \frac{n_0^2/\chi_{\perp}}{\omega_{\text{AFM}}^2({\bm k}) - z^2}\ .
\label{eq:A.5a}
\ee
Here $n_0$ is the staggered magnetization amplitude, and $\chi_{\perp}$ is the transverse homogeneous 
susceptibility (which is finite). The resonance frequency
\be
\omega_{\text{AFM}}({\bm k}) = c\,\vert{\bm k}\vert
\label{eq:A.5b}
\ee
\ese
shows the linear dispersion relation for antiferromagnetic magnons, with $c$ the spin-wave velocity. 

Note that $\chi_{\text{AFM}}$ has the same form as the displacement susceptibility $\chi_u$ in Eq.~(\ref{eq:A.4a}).
$\chi_{\text{AFM}}({\bm k},z)$ thus is the Goldstone susceptibility, and Eqs.~(\ref{eqs:A.2}) apply for
intra- and interband scattering, respectively. Antiferromagnons couple electrons in the same sub-band,
and therefore intraband scattering dominates in the $T\to 0$ limit.

\subsubsection{Helimagnons}
\label{app:A.1c}

In helical magnets the relevant susceptibility is proportional to a phase susceptibility of the form\cite{Belitz_Kirkpatrick_Rosch_2006b} 
\bse
\label{eqs:A.6}
\be
\chi_{\text{HM}}({\bm k},z) = \frac{m_0 D q^2}{\omega_{\text{HM}}^2({\bm k}) - z^2}\ .
\label{eq:A.6a}
\ee
Here $m_0$ is the amplitude of the helically modulated magnetization, and $D$ is a spin stiffness coefficient.
For wave numbers smaller than the pitch wave number $q$ the resonance frequency is anisotropic\cite{Belitz_Kirkpatrick_Rosch_2006a}
\be
\omega_{\text{HM}}({\bm k}) = D\sqrt{q^2 k_z^2 + {\bm k}_{\perp}^4}
\label{eq:A.6b}
\ee
\ese
with $k_z$ and ${\bm k}_{\perp}$ the components of ${\bm k}$ parallel and perpendicular to the pitch wave vector ${\bm q}$, respectively.
For wave numbers larger than $q$ the resonance frequency crosses over to the ferromagnetic one given in Sec.~\ref{app:A.1d} below.
The Fermi surface is split, and the same considerations about intraband versus interband scattering as in the antiferromagnetic
case apply. 

We note that helical magnets also support columnar phases of skyrmionic type,\cite{Muehlbauer_et_al_2009} where the relevant
resonance frequency is similar to $\omega_{\text{HM}}$, but with the roles of $k_z$ and ${\bm k}_{\perp}$ 
interchanged.\cite{Kirkpatrick_Belitz_2010, Petrova_Tchernyshyov_2011} This case can be treated in complete analogy to the
 helimagnon case.

\subsubsection{Ferromagnons}
\label{app:A.1d}

%
%
The ferromagnetic case is different from all others in several respects. First, the
resonance frequency is quadratic in ${\bm k}$,
\be
\omega_{\text{FM}}({\bm k}) = D{\bm k}^2\ .
\label{eq:A.7}
\ee
Second, the ferromagnons do not couple electrons in the same sub-band, so only interband scattering is present
and the coupling does not come with a gradient squared, see the discussion around Eqs.~(\ref{eqs:A.2}). 
The sub-bands are characterized by the spin projection $\sigma = \uparrow,\downarrow \equiv \pm$, and the
potential $V$ from Eq.~(\ref{eq:A.1a}), with the spin labels written explicitly, has the form
\bse
\label{eqs:A.8}
\be
V_{\sigma\sigma'}({\bm k},z) = \frac{v_0}{\NF^2}\,(1 - \delta_{\sigma\sigma'})\,\chi_{\sigma'}({\bm k},z)
\label{eq:A.8a}
\ee
which explicitly shows the interband nature of the scattering process.
The susceptibility $\chi_{\sigma}$ describes a circularly polarized ferromagnon with amplitude $m_0$,
\be
\chi_{\pm}({\bm k},z) = \frac{m_0}{\omega_{\text{FM}}({\bm k}) \pm z}\ .
\label{eq:A.8b}
\ee
\ese
The linear combinations $\chi_{+} \pm \chi_{-}$ yield the matrix elements of the transverse magnetic susceptibility,\cite{Forster_1975, Chaikin_Lubensky_1995}
which have the form of Eq.~(\ref{eq:A.1b}). 

The resulting contributions to the transport coefficients will be discussed in Paper II.

\subsection{Effective potentials}
\label{app:A.2}

In an approach to the transport problem via the Kubo formula, the effective potential $V$ enters via its spectrum $V''$ averaged 
over energy shells at fixed distances from the chemical potential.\cite{Mahan_2000} For the intraband case this average has the form
\be
\frac{1}{\NF^2 V^2} \sum_{{\bm k},{\bm p}} V''({\bm k}-{\bm p},u)\,\delta(\xi_{\bm k} - \epsilon)\,\delta(\xi_{\bm p} - \epsilon - u)
\label{eq:A.9}
\ee
The phonon and antiferromagnon cases share the same
spectrum of the effective potential, viz.,
\be
V''({\bm k},u) = \frac{g_0}{\NF}\,\frac{c^2 k^2}{u}\, \left[\delta(u - ck) + \delta(u + ck)\right]
\label{eq:A.10}
\ee
where the dimensionless constant $g_0$ is determined by Eqs.~(\ref{eq:A.3a}) or (\ref{eq:A.4a}) for phonons, and by (\ref{eq:A.5a})
for antiferromagnons.

The energy variable $\epsilon$ scales as the temperature by virtue of the weight function $w$ in Eqs.~(\ref{eqs:2.15}).
The $\epsilon$ in the arguments of the $\delta$-functions therefore leads only to effects of $O(T/\epsilonF)$ that do not
contribute to the leading low-$T$ behavior.\cite{thermopower_footnote_2} The $u$ in the argument of the second $\delta$-function
leads to the kernel $K_1$, and in addition to corrections to the limits of the wave-number integration that again do not
contribute to the leading low-$T$ behavior. The effective potential that determines the kernel $K_0$ is thus
\bea
{\bar V}''(u) &=& \frac{1}{\NF^2 V^2} \sum_{{\bm k},{\bm p}} V''({\bm k}-{\bm p},u)\,\delta(\xi_{\bm k})\,\delta(\xi_{\bm p})
\nonumber\\
&\approx& \frac{g_0}{\NF}\, (u/T_1)^2 \sgn(u)\,\Theta(T_1 - \vert u\vert)
\nonumber\\
&=& \frac{1}{\NF}\,V_0(u)
\label{eq:A.11}
\eea
where $T_1 = 2c\kF$ and $V_0$ is the function from Eq.~(\ref{eq:2.13b}).

In helimagnets, intraband scattering also dominates, but the anisotropic excitation spectrum makes the wave number
integrals more complicated. The analogous effective potential for interband scattering, which is relevant for ferromagnets, will be discussed in Paper II.

\section{Relaxation rates}
\label{app:B}

In this appendix we discuss the relaxation rates defined in Eqs.~(\ref{eqs:2.19}).

\begin{figure}[t]
\includegraphics[width=6cm]{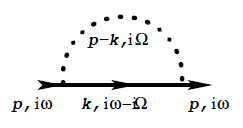}
\caption{Exchange contribution to the electronic self energy.}
\label{fig:1}
\end{figure}

The single-particle relaxation rate $\Gamma_0$ is usually defined as the imaginary part of the electronic self energy.
To lowest order in the spectrum of the effective potential $V''({\bm k},\omega)$, Eq.~(\ref{eq:A.10}), it is given by
the diagram shown in Fig.~\ref{fig:1}.
In terms of integrals this diagram reads
\bea
\Gamma_0(\epsilon) &=& \int_{-\infty}^{\infty} du\,\left[n_0\left(\frac{u}{T}\right) + f_0\left(\frac{u+\epsilon}{T}\right)\right] \frac{1}{V}\sum_{\bm k} V''({\bm k},u)\nonumber\\
&& \hskip 1pt \times \frac{1}{V}\sum_{\bm p} \delta(\xi({\bm p}+{\bm k})-\epsilon-u)\,\delta(\xi({\bm p})-\epsilon)\ .
\nonumber\\
\label{eq:B.1}
\eea
An inspection shows that this is identical with the rate $\Gamma_0$ defined in Eqs.~(\ref{eq:2.1a}). 

For the scaling function $\gamma_0$, Eq.~(\ref{eq:2.19b}), this yields
\bse
\label{eqs:B.2}
\be
\gamma_0(x) = 2\,g_0\left[2\,\zeta(3) - \text{Li}_3(-e^x) - \text{Li}_3(-e^{-x})\right]
\label{eq:B.2a}
\ee
with $\zeta$ the Riemann zeta function, and $\text{Li}_n$ the polylogarithm. For the other two
scaling functions in Eqs.~(\ref{eqs:2.18}) we have
\bea
\gamma_1(x) &=& 6\,g_0\left(\text{Li}_4(-e^x) - \text{Li}_4(-e^{-x})\right)\ ,
\label{eq:B.2b}\\
\gamma_2(x) &=& 24\,g_0\left[2\,\zeta(5) - \text{Li}_5(-e^x) - \text{Li}_5(-e^{-x}\right]\ .\qquad
\label{eq:B.2c}
\eea
\ese

The single-particle relaxation rate for a quasiparticle on the Fermi surface shows the well-known $T^3$ behavior:
\be
\Gamma_0(\epsilon=0) = 7\, \zeta(3) g_0\,T^3/T_1^2\ .
\label{eq:B.3}
\ee
It is illustrative to explain the origin of the $T^3$ behavior in a scaling sense. $u$ scales as $T$, $u\sim T$,
and $k\sim u \sim T$ due to the linear phonon spectrum. The integration measures for the $u$ and ${\bm k}$
integrations in Eq.~(\ref{eq:B.1}) thus scale as $du \sim T$, $dk\,k^2 \sim T^3$, and the spectrum of the potential,
Eq.~(\ref{eq:A.10}), scales as $V'' \sim k^2/u^2 \sim T^0$. Finally, the convolution integral over the product
of $\delta$-functions produces a factor of $1/k \sim T^{-1}$, so $\Gamma_0 \sim T^{1+3+0-1} = T^3$.

Due to the $\epsilon/T$ scaling of $\Gamma_0$, the $T$-dependence
of the on-shell rate, Eq.~(\ref{eq:B.3}), is the same as the one of the rate averaged over $\epsilon$ with
the weight function $w(\epsilon)$ from Eq.~(\ref{eq:2.16a}),
\bse
\label{eqs:B.4}
\be
\langle\Gamma_0\rangle_w \equiv \frac{1}{T} \int d\epsilon\,w(\epsilon)\,\Gamma_0(\epsilon) =12\,\zeta(3)\,g_0\,T^3/T_1^2\ .
\label{eq:B.4a}
\ee
Similarly, we obtain 
\bea
\langle\epsilon\Gamma_1\rangle_w &=& 60\,\zeta(5)\, g_0\,T^5/T_1^2\epsilonF\ ,
\label{eq:B.4b}\\
\langle\Gamma_2\rangle_w &=& 120\,\zeta(5)\,g_0\,T^5/T_1^4\ .
\label{eq:B.4c}
\eea
\ese
In performing these integrals, it is helpful to do the $\epsilon$-integration first.

\section{Results for helimagnons}
\label{app:C}

The transport coefficients due to helimagnon scattering can be derived in analogy to the phonon and
antiferromagnon cases by using the susceptibility and resonance frequency given in Eqs.~(\ref{eq:A.6a}) and
(\ref{eq:A.6b}), respectively, in the effective potential. This calculation establishes that the result for the electrical
conductivity first obtained in Ref.~\onlinecite{Belitz_Kirkpatrick_Rosch_2006b} 
(see also Refs.~\onlinecite{Kirkpatrick_Belitz_Saha_2008a, Kirkpatrick_Belitz_Saha_2008b}) is exact. Here we recall how this
result comes about in a scaling sense, and give the corresponding temperature dependences of the thermal
conductivity and the thermopower. 

We recall the scaling arguments for the single-particle rate $\Gamma_0$, which scales as the collision
operator $\Lambda_0$, in the phonon case as given after Eq.~(\ref{eq:B.3}), and consider the modifications
for the helimagnon case. The anisotropy of the resonance frequency, Eq.~(\ref{eq:A.6b}), implies that $k_z$ 
and ${\bm k}_{\perp}$ scale differently with temperature, viz., $k_z \sim k_{\perp}^2 \sim T$. The 
${\bm k}$-integration measure thus scales as $dk_z\,d{\bm k}_{\perp}^2 \sim T^2$. Similarly, the gradient-squared 
coupling from Eq.~(\ref{eq:A.2a}) scales as  $k_{\perp}^2 \sim T$, and hence $V'' \sim T^{-1}$.
Finally, the ${\bm p}$-integration still produces a factor of $1/k$, which now scales as $1/k \sim 1/k_{\perp} \sim T^{-1/2}$. 
Together, we have $\Gamma_0 \sim T^{1+2-1-1/2} = T^{3/2}$. Collecting powers of the pitch wave number $q$
and the exchange splitting $\lambda$ one finally obtains\cite{Belitz_Kirkpatrick_Rosch_2006b} 
\be
\Gamma_0(\epsilon=0) \propto \lambda\,\left(\frac{q}{\kF}\right)^6 \left(\frac{\epsilonF}{\lambda}\right)^2 \left(\frac{T}{D q^2}\right)^{3/2}\ .
\label{eq:C.1}
\ee
This result holds for an electronic single-particle spectrum $\epsilon_{\bm k}$ with cubic symmetry. For nearly free
electrons the gradient-squared coupling is effectively a $k_z^2 \sim T^2$,
which leads to $\Gamma_0 \sim T^{5/2}$.\cite{Belitz_Kirkpatrick_Rosch_2006b}

For the function $V_0$ that determines the kernel $K_0$ via Eq.~(\ref{eq:2.13a}) this implies
\bea
V_0(u) \propto (\vert u\vert/T_1)^{1/2} \sgn(u)\,\Theta(T_1 - \vert u\vert)\ .
\label{eq:C.2}\\
\nonumber
\eea
The magnetic Debye temperature in this context is $T_1 = D\kF^2$ with $D$ the spin stiffness coefficient
from Eq.~(\ref{eq:A.6a}). The kernel $K_2$ is determined by an effective potential that is given by a
Fermi-surface average analogous to Eq.~(\ref{eq:A.11}), but with an additional factor of $({\bm k}-{\bm p})^2/\kF^2$
in the integrand that scales as $({\bm k}-{\bm p})_{\perp}^2 \sim \vert u \vert$. As a result, Eq.~(\ref{eq:2.14d}) gets
replaced by
\be
K_2(\epsilon,u) = \frac{\vert u - \epsilon\vert}{2 T_1}\,K_0(\epsilon,u)\ ,
\label{eq:C.3}
\ee
while $K_1$ is still given by Eq.~(\ref{eq:2.14c}). $\Lambda_1$ and $\Lambda_2$ thus both scale the same way, 
$\Lambda_1 \sim \Lambda_2 \sim T^{5/2}$. The skew-adjointness of $\Lambda_1$ then implies that it does
not contribute to the leading temperature dependence of the electrical resistivity. The latter is determined by
$\Lambda_2$ alone, and the result is given by Eq.~(\ref{eq:C.1}) with an additional factor of $T/T_1$,\cite{Belitz_Kirkpatrick_Rosch_2006b}
\be
\rho \propto \lambda \left(\frac{q}{\kF}\right)^8 \left(\frac{\epsilonF}{\lambda}\right)^2 \left(\frac{T}{Dq^2}\right)^{5/2}\ .
\label{eq:C.4}
\ee 

The same result can be obtained by adapting the projector method from Sec.~\ref{subsec:III.B}: Since $\Gamma_2$ now
scales as $\langle\Gamma_2\rangle_w \sim T\,\Gamma_0 \sim T^{5/2}$, whereas $\langle\epsilon\Gamma_1\rangle_w \sim T^2 \Gamma_0 \sim T^{7/2}$,
$\phi_0$ is given by 
\bse
\label{eqs:C.5}
\be
\phi_0 = 1/\langle\Gamma_2\rangle_w\ . 
\label{eq:C.5a}
\ee
To leading order, $\varphi_0$ is still given by 
\be
\vert\varphi_0\rangle = \phi_0 \vert 1\rangle\ ,
\label{eq:C.5b}
\ee
\ese
and Eq.~(\ref{eq:3.2a}) yields Eq.~(\ref{eq:C.4}) for the electrical resistivity.

The thermal transport behavior is also most easily analyzed by means of the projector method. 
If we adjust the procedure in Sec.~\ref{subsec:III.B} for the fact that $\Lambda_1$ and $\Lambda_2$ now
both scale the same way, we obtain, to leading order,
\be
P\vert\varphi_1\rangle \propto \frac{\langle\epsilon^2\rangle_w}{\langle\Gamma_2\rangle_w}\,\vert 1\rangle\ .
\label{eq:C.6}
\ee
For the thermopower, Eq.~(\ref{eq:3.2b}) then yields
\be
S \propto T/\epsilonF\ ,
\label{eq:C.7}
\ee
as for phonons and antiferromagnons. However, in contrast to the phonon case, $P_{\perp}\vert\varphi_1\rangle$
and $P \vert\varphi_1\rangle$ now scale the same way. Consequently, the leading contribution to $P_{\perp}\vert\varphi_1\rangle$
is given by the kinetic part alone,
\be
P_{\perp}\vert\varphi_1\rangle = -\vert h\rangle\ ,
\label{eq:C.8}
\ee
with $\vert h\rangle$ from Eqs.~(\ref{eqs:3.21}), while the hydrodynamic part is subleading. The scaling arguments
given in Sec.~\ref{subsec:IV.C} still apply,  and the dominant contribution to the heat diffusivity thus is, apart from a numerical prefactor,
\be
\sigma_h/T \propto \frac{1}{\lambda} \left(\frac{\kF}{q}\right)^6 \left(\frac{\lambda}{\epsilonF}\right)^2 \left(\frac{D q^2}{T}\right)^{3/2}\ .
\label{eq:C.9}
\ee


\begin{thebibliography}{28}
\bibitem[{the({\natexlab{a}})}]{thermopower_footnote}
\bibinfo{note}{In particular, the thermopower can be expressed in terms of an
  integral over $\epsilon\,\varphi_0$ rather than the integral over $\varphi_1$
  in Eq.~(\ref{eq:2.15b}). In order to show the equivalence of these two
  expressions one needs to keep the terms of $O(\epsilon/\epsilonF)$ that we
  have neglected.}
\bibitem[{LR_()}]{LR_footnote}
\bibinfo{note}{In order to avoid cumbersome notation we always denote right
  eigenvectors by $\vert e_n\rangle$ and left eigenvectors by $\langle
  e_n\vert$. It is important to remember that these two vectors represent
  different functions. In particular, $\langle f\vert e_n\rangle \neq \langle
  e_n\vert f\rangle$ for a given function $f$. A more precise, if more
  complicated, notation would be to denote the right eigenfunctions by $e_n^R$,
  and the left ones by $e_n^L$, in which case $\langle f\vert e_n^R\rangle =
  \langle e_n^R\vert f\rangle \neq \langle e_n^L\vert f\rangle$.}
\bibitem[{the({\natexlab{b}})}]{thermopower_footnote_2}
\bibinfo{note}{This statement does not hold for an alternative expression for
  the thermopower that is equivalent to Eq.~(\ref{eq:2.15b}), see
  Ref.~\onlinecite{thermopower_footnote}.}
\end{thebibliography}

\end{document}